\documentclass[10pt]{iopart}
\expandafter\let\csname equation*\endcsname\relax
\expandafter\let\csname endequation*\endcsname\relax
\usepackage{amsmath}
\usepackage{iopams}
\usepackage[dvips]{graphicx}
\usepackage{subfig}

\newcommand{\Psiscat}[1]{\Psi_{#1}^{(-)}}
\newcommand{\Psineut}[1]{\Phi_{#1}^{N}}
\newcommand{\psik}[2]{\psi_{k#1}^{(N)\mathrm{#2}}}
\newcommand{\psitg}[2]{\Phi_{#1}^{(N-1)\mathrm{#2}}}

\newcommand{\csflsq}[2]{\chi_{#1}^{(N)\mathrm{#2}}}

\newcommand{\radialFnKmat}{\boldsymbol{F}}

\newcommand{\radialFnSmat}[1]{\boldsymbol{F}^{#1}}
\newcommand{\radialFnPSmat}[1]{\boldsymbol{F}'^{#1}}
\newcommand{\ctorb}[1]{\eta_{#1}}

\newcommand{\asymop}{\mathcal{A}}
\newcommand{\opdipole}{\boldsymbol{d}}
\newcommand{\braopket}[3]{\langle {#1}|{#2}|{#3}\rangle}
\newcommand{\braket}[2]{\langle {#1}|{#2}\rangle}

\newcommand{\matrx}[1]{\mathbf{#1}}

\newcommand{\AkCoeffMat}[1]{\boldsymbol{A}^{#1}}

\newcommand{\rotmat}[3]{D^{#1}_{#2 #3}}
\newcommand{\rerotmat}[3]{\Delta^{#1}_{#2 #3}}
\newcommand{\spharm}[2]{Y_{#1,#2}}
\newcommand{\teharm}[2]{S_{#1,#2}}
\newcommand{\euler}{\alpha,\beta,\gamma}

\newcommand{\MTXrotmat}[1]{\boldsymbol{D}^{#1}}
\newcommand{\MTXrerotmat}[1]{\boldsymbol{\Delta}^{#1}}
\newcommand{\coulombphase}[1]{\sigma_{#1}}

\newcommand{\kfinalMOL}{\boldsymbol{k}_{f}}
\newcommand{\kdirMOL}{\boldsymbol{\hat k}_{f} }
\newcommand{\kfinalLAB}{\boldsymbol{k}_{f}'}
\newcommand{\kdirLAB}{\boldsymbol{\hat k}_{f}'}

\newcommand{\xtot}{x_1,...,x_{{N-1}},x_{N}}
\newcommand{\xtarg}{x_1,...,x_{{N-1}}}

\mathchardef\mhyphen="2D

\begin{document}
\title[PAD from aligned molecules using 
the R-matrix method]{Photoelectron angular distributions from aligned molecules using 
the R-matrix method }
\author{Alex G Harvey, Danilo S Brambila, Felipe Morales and Olga Smirnova}
\address{Max-Born-Institut, Max-Born-Strasse 2A, D-12489 Berlin, Germany}
\ead{harvey@mbi-berlin.de}

\begin{abstract}
We present a new extension of the UKRmol 
electron-molecule scattering code suite, which allows one to compute \textit{ab initio} 
photoionization and photorecombination amplitudes for complex molecules, 
resolved both on the molecular alignment (orientation) and the emission angle 
and energy of the photoelectron. 
We illustrate our approach using CO$_2$ as an example, and 
analyze the importance of multi-channel effects by performing our 
calculations at different, increasing levels of complexity. 
We benchmark our method by comparing the results of our calculations with 
experimental data and with theoretical calculations available in the literature. 
\end{abstract}
\maketitle

\section{Introduction}

The goal of attosecond spectroscopy is to observe non-equilibrium 
multi-electron dynamics on its natural, attosecond, time-scale. 
In the family of recently developed approaches, including high harmonic 
generation (HHG) spectroscopy \cite{baker06,smirnova09,haessler10}, the 
attosecond streak camera \cite{schultze10}, the reconstruction of attosecond 
bursts by interference of two-photon transitions (RABBITT) method 
\cite{paul01,kluender11}, laser-induced electron diffraction 
\cite{spanner04,meckel08,blaga12}, attosecond resolution comes hand 
in hand with the application of intense IR or XUV pulses and usually involves 
ionization of the target. 
The attosecond dynamics are often mapped onto, and have to be read out from, 
the atomic or molecular continuum. 
These attosecond dynamics can be decoded from experimental observables using 
semi-classical methods \cite{ivanov96,jin09,ivanov11}, 
which often allow one to break the full process into separate, coherent steps, 
e.g. ionization, continuum dynamics, recombination or scattering. 
Decoding such attosecond dynamics requires the development of theoretical 
methods capable of describing angle and energy resolved 
photoionization/photorecombination dipoles and scattering amplitudes. 
Accurate treatment of the molecular continuum is crucial for pushing 
attosecond spectroscopy to complex molecules. 
Most attosecond experiments are done with aligned molecules, but angular 
and energy resolved dipoles are often required even in the case of unaligned 
or weakly aligned targets to account for the coherence between different 
pathways associated with different alignment angles. 
For example, coherent addition of light emitted during the recombination of an 
electron with its parent molecular ion, for different molecular orientations, 
is crucial for HHG spectroscopy. 
Significant advances in HHG spectroscopy are associated with the application 
of the Schwinger variational method for calculating 
photoionization/photorecombination dipoles \cite{le09}. 
Here we present a new development based on the extension of the UKRmol 
\cite{carr12} electron-molecule scattering codes, which allows one to compute 
dipole matrix elements, and photoionization and recombination cross-sections, 
resolved in alignment, emission angle and energy, for complex molecules.  
As an example, we consider the application of the new codes to the CO$_2$ 
molecule.

We note that there have been relatively few applications of the R-matrix 
approach to molecular photoionization 
\cite{tennyson87,tennyson87a,colgan01,tashiro10} and they have primarily 
looked at orientationally averaged observables, the exception being the 
single channel, finite element, R-matrix code, FERM3D \cite{tonzani07}.

\section{Theory}
\subsection{Overview}
Initially proposed by Wigner and Eisenbud \cite{wigner47} in the 1940s for the 
characterization of resonant nuclear scattering, the R-matrix method has 
undergone, over the years, significant adaptation and development to treat 
electron-atom and electron-molecule interactions. We will present here just 
those elements of R-matrix theory required to describe our new development. 
For comprehensive discussion of R-matrix techniques and applications, as 
applied to electron and photon induced processes in atomic and molecular 
physics, the recent review \cite{tennyson10}, and book \cite{burke11} are 
excellent starting points.

The R-matrix technique is a method of solving the, inherently multi-channel, 
electron-(photon)-molecule collision problem within the close coupling 
approximation. 
The power of the R-matrix approach lies in the division of the configuration 
space of the molecule into separate regions; this division allows one to apply 
the appropriate approximation and optimal computational technique in each 
region.

The usual division is into an inner region, close to the molecule, where the 
non-local electron-electron exchange and correlation interactions are 
important, and must be accounted for; an outer region, where the continuum 
electron is distinguishable from the bound electrons,  non-local exchange and 
correlation are negligible, and the problem reduces to the ejected electron 
scattering in the long-range multi-pole potential of the parent molecule; 
and an asymptotic region, where the long-range potential is weak, and the 
solution is well represented by an asymptotic expansion which satisfies the 
physical boundary conditions.

In the inner region the continuum is discretized, allowing the use of basis 
set methods adapted from quantum chemistry. 
In the outer region numerical integration techniques are used to propagate 
the inner region solutions from the boundary to the asymptotic region, where 
matching to the asymptotic expansion applies the appropriate physical 
boundary conditions.
 
\subsection{Dipoles and cross sections}

UKRmol calculations are performed in the molecular frame, so in the 
following discussion we start from the molecular frame description of 
photoionization and, towards the end, transform to the laboratory frame. 
Along the way we will indicate the new (in contrast to a scattering 
calculation) quantities that we require.

In the theoretical description of single photon ionization \cite{bethe57}, 
within the length gauge dipole approximation, the molecular frame 
photo-electron angular distribution can be expressed as,
\begin{equation} \label{eqn_mfpad}
   \frac{\mathrm{d}\sigma_{fi}}{\mathrm{d}\kfinalMOL}=4\pi^2\alpha a_0^2 
   \omega|\boldsymbol{d}_{fi}(\kfinalMOL)\cdot \boldsymbol{\hat \epsilon}|^2,
\end{equation}
where $\alpha$ is the fine structure constant, $a_0$ is the Bohr radius, 
$\omega$ is the photon energy in atomic units, and $\boldsymbol{\hat \epsilon}$ 
is the polarization direction of the incident photon in the molecular frame. 
The molecular frame transition dipole, $\boldsymbol{d}_{fi}(\kfinalMOL)$, 
between an initial bound state, $\Psineut{i}$, and a final continuum state, 
$\Psiscat{f\kfinalMOL}$, of the molecule is,
\begin{equation} \label{eqn_dipole}
   \boldsymbol{d}_{fi}(\kfinalMOL)=\braopket{\Psiscat{f\kfinalMOL}}{\opdipole}{\Psineut{i}},
\end{equation}
where $\opdipole$ is the dipole operator, the momentum of the ejected electron 
is $\kfinalMOL$, and the ion is left in the state indexed by $f$. 
We use the length gauge form of the dipole operator, which can be written in 
spherical vector form as,
\begin{equation} \label{eqn_dipole_operator}
d_q=\left (\frac{4\pi}{3} \right )^{1/2} \sum_{i=1}^{N}r_i \spharm{1}{q}(\boldsymbol{\hat r}_i),
\end{equation}
where $\boldsymbol{\hat r}_i$ is the coordinate of the $i$-th electron and 
$\spharm{1}{q}(\boldsymbol{\hat r}_i)$ is a spherical harmonic. 
The $q=\pm 1$ components correspond to circular polarization and the $q=0$ 
component to linear polarization.

If we only consider initial states that fit in the inner region, the integral 
above can be restricted to the inner region and we can expand both the 
initial and final state in terms of the energy independent 
inner region solutions $\psik{}{}$.
\begin{eqnarray} 
\Psiscat{f\kfinalMOL} &=& \sum_k A_{fk}(\kfinalMOL)\psik{}{} \\ \label{eqn_expanded_scat}
\Psineut{i} &=&  \sum_k B_{ik}\psik{}{}       \label{eqn_expanded_bound}
\end{eqnarray} 
The transition dipole becomes
\begin{eqnarray} \label{eqn_rmat_photodipole_momentum}
\boldsymbol{d}_{fi}(\kfinalMOL)=\sum_{kk'}A^{*}_{fk}(\kfinalMOL)\braopket{\psik{}{}}{\opdipole}{\psik{'}{}}B_{ik'},
\end{eqnarray} 
The R-matrix codes work in the angular momentum basis for the ejected 
electron, in a partial wave expansion, as the expansion 
often converges for low values of $l$. In this basis, eq. 
\ref{eqn_rmat_photodipole_momentum} becomes,
\begin{eqnarray} \label{eqn_rmat_photodipole_angular_momentum}
\boldsymbol{d}_{fi}(\kfinalMOL)&=&\sum_{kk'}\sum_{l_f m_f}i^{-l_f}{e^{i\coulombphase{l_f}}}\spharm{l_f}{m_f}(\kdirMOL)(S_f M_{S_f} \tfrac{1}{2} m_{s_f}|S M_S) \nonumber \\ 
&&A^{*}_{fl_fm_f,k}(E) \braopket{\psik{}{}}{\opdipole}{\psik{'}{}}B_{ik'} ,
\end{eqnarray} 
here $\coulombphase{l_f}=\arg \Gamma(l_f+1+i\eta_f)$ is the Coulomb phase,
with $\eta_f=-\frac{Z-(N-1)}{k_f}$, where $Z-(N-1)$ is the residual charge on the 
the ion.
The Clebsch-Gordan coefficient is due to spin coupling of the continuum 
electron and the ion.
The partial wave dipole is defined as follows.
\begin{eqnarray}\label{eqn_partial_wave_dipole_mol}
\boldsymbol{d}_{fl_fm_f,i}(E)=\sum_{kk'} A^{*}_{fl_fm_f,k}(E)\braopket{\psik{}{}}{\opdipole}{\psik{'}{}}B_{ik'}
\end{eqnarray} 
Clearly, to calculate transition dipoles and photoionization/recombination 
observables we need the expansion coefficients for the initial and final 
states in the inner region and transition dipoles between the inner region 
states. 
In contrast, only the expansion coefficients in terms of the 
asymptotic solutions are required to obtain scattering observables.

To connect to the lab frame observables, we introduce Euler angles $\euler$, 
which define the rotation of molecule from the lab frame, and the 
associated Wigner rotation matrices (see, for example \cite{brink93}), 
$\MTXrotmat{l}(\euler)$. In the lab frame 
eq.~\ref{eqn_partial_wave_dipole_mol} becomes,
\begin{eqnarray}\label{eqn_partial_wave_dipole_lab}
\boldsymbol{d}_{fi}'(\kfinalLAB)&=&\sum_{l_f m_f' m_f}i^{-l_f}{e^{i\coulombphase{l_f}}}\spharm{l_f}{m_f'}(\kdirLAB)(S_f M_{S_f} \tfrac{1}{2} m_{s_f}|S M_S) \nonumber \\ 
&&\rotmat{l}{m_f'}{m_f} \boldsymbol{d}_{fl_fm_f,i}(E)\MTXrotmat{1\dagger},
\end{eqnarray} 
where the primed variables indicate lab frame quantities.
If the target molecule spins are unpolarized and the final state spins are not 
measured, one must average over the initial and sum over the final spin 
components.
\begin{equation} \label{eqn_pad_spin_averaged}
   \frac{\mathrm{d}\sigma_{fi}}{\mathrm{d}\kfinalMOL}=\frac{4\pi^2\alpha a_0^2\omega}{2S+1} \sum_{M_S,M_{S_f}, m_{s_f}}
   |\boldsymbol{d}_{fi}(\kfinalMOL)\cdot \boldsymbol{\hat \epsilon}|^2,
\end{equation}
Finally, in addition to angular distributions from aligned molecules, we will 
present orientationally averaged partial photoionization cross sections for 
comparison to existing theory and experiment. 
With the aid of angular momentum algebra (see, for example, \cite{burke82}) 
the orientationally averaged partial photoionization cross section takes the 
following convenient form,
\begin{eqnarray}\label{eqn_orient_av_xsec}
    \sigma_{fi}(E)=\frac{4}{3}\pi^2\alpha a_0^2 
   \omega\sum_{q l_f m_f}|d_{q,fl_fm_f,i}(E)|^2.
\end{eqnarray}
photoionization and recombination observables are produced by the new code 
module, DIPELM, which takes the expansion coefficients and inner region 
dipoles as input.
In the next section, we describe the R-matrix approach to calculating the 
scattering wavefunction,  $\Psiscat{f\kfinalMOL}$. 

\subsection{Scattering states}

The R-matrix method, as implemented by the UKRmol code suite, is a 
close-coupling approach to solving the scattering problem (within the fixed 
nuclei approximation), where the scattering wavefunction is expanded in 
terms of target states, with expansion coefficients dependent on the 
coordinate of the continuum electron. 

\subsubsection{Inner region}
In the inner region, the (discretized) continuum is represented by a set of 
continuum orbitals, $\ctorb{m_{f}}$, and the close-coupling expansion takes 
the form   
\begin{eqnarray}\label{eqn_closecoup_inner}
\psik{}{}&=&\asymop \sum_{f m_{f}} 
           a_{k f m_{f}}\psitg{f}{}(\xtarg)\ctorb{m_{f}}(x_{N}) \nonumber \\
          &+& \sum_{p} b_{kp} \csflsq{p}{}(\xtot).
\end{eqnarray}
The continuum orbitals are built from a set of Gaussian type orbitals (GTO) 
fitted to Coulomb or Bessel functions \cite{faure02}. 
Molecular orbitals are constructed from a second set of GTO and used to 
create the target states, $\psitg{f}{}$, using configuration interaction. 
$\csflsq{p}{}$ are configurations created by placing the continuum 
orbital into a bound orbital and are needed to describe short-range 
correlation lost due to orthogonalization of bound and continuum orbitals. 
Exchange is treated rigorously by anti-symmetrization (the operator 
$\asymop$).
The coefficients, $a_{k f m_{f}}$ and  $b_{kp}$, are found by diagonalising the 
full electronic Hamiltonian restricted to the inner region \cite{tennyson96}. 
The end result of this procedure is a flexible basis, $\psik{}{}$, with which 
to represent the N-electron wavefunction, for both bound and continuum states, 
in the inner region. 

To connect to the outer region, the radial part of the inner region 
wavefunctions, evaluated on the R-matrix boundary is needed (only the 
continuum orbitals are non-zero on the boundary). These are known as the 
boundary amplitudes, $ w_{i k}$, and are constructed by projection 
on to the channel functions as follows,
\begin{eqnarray}
 w_{ik}(a)=\braket{\psitg{i}{}\spharm{l_i}{m_i}}{\psik{}{}}.
\end{eqnarray}
The spherical harmonics, $\spharm{l_i}{m_i}$, are defined in the molecular 
frame and we note that the continuum is spin coupled to the target. 
The partial wave expansion of the continuum implicit in the above leads to 
each target state being associated with a number of degenerate partial wave 
channels.
We denote the total number of these channels as $n$, with $n_o$ energetically 
open and $n_c$ closed.
The single index, $i$, on the boundary amplitudes now indexes these channels.
The $n\times n$ R-matrix is then constructed,
\begin{eqnarray}
\matrx{R}(a)=  \frac{1}{2} \matrx{w}(a)  \left[ \matrx{E_k}-E \right]^{-1} \matrx{w}^T(a),
\end{eqnarray}
where $\left[ \matrx{E_k}-E \right]^{-1}$ is a diagonal matrix with elements 
$\delta_{kk'}(E_k-E)^{-1}$. 
$E_k$ are the eigenenergies of the $\psik{}{}$ and are known as the R-matrix 
poles.
If $\Psi_{j}^{}$ is a solution of the full Hamiltonian, then projecting out 
the channel functions, and evaluating at $r=a$ as before gives us,
\begin{eqnarray}
 F_{ij}(a)=\braket{\psitg{i}{}\spharm{l_i}{m_i}}{\Psi_{j}^{}},
\end{eqnarray}
the R-matrix (in its simplest form) then relates $\matrx{F}$ to its 
derivative,
\begin{eqnarray}
\matrx{F}(a)=\matrx{R}(a)\matrx{F'}(a).
\end{eqnarray}
\subsubsection{Outer and asymptotic region}

The close coupling expansion in the outer region takes a simpler form,
\begin{eqnarray}\label{eqn_closecoup_outer}
\Psi_{j}=\sum_{i} 
           \psitg{i}{}(\xtarg;\sigma_{N+1})\spharm{l_i}{m_i}(\hat r_{N+1})r_{N+1}^{-1}F_{ij}(r_{N+1})
\end{eqnarray}
where the summation over $i$ is a summation over the partial wave channels.
Substituting into the Schrödinger equation and projecting out the channel 
functions give the reduced radial equations,
\begin{eqnarray}
 \left[ \frac{\mathrm{d}^2}{\mathrm{d}r^2} - \frac{l_i(l_i+1)}{r^2} +\frac{2(Z-(N-1))}{r} + k_i^2 \right ]F_{ij}=2\sum_{i'=1}^{n} V_{ii'}F_{i'j}.
\end{eqnarray}
Of the $2n$ linearly independent solutions to the reduced radial equations, 
$n$ are divergent at the origin and $n_c$ are divergent asymptotically, 
and thus, are physically inadmissible. 
This leaves $n_o$ independent solutions.
$\radialFnKmat$ has standing wave asymptotic boundary 
conditions,
\begin{eqnarray}
\radialFnKmat \sim \matrx{k}^{-1/2}[\matrx{S}+\matrx{CK}],
\end{eqnarray}
where $\matrx{S}$ is the matrix of fundamental solutions with sine 
like asymptotic behaviour $\matrx{C}$ is cosine like for open channels and 
exponentially decaying for closed channels. Computationally, it is useful to 
apply standing wave boundary conditions and solve for the K-matrix as it 
avoids the need for complex numbers in the calculation until the end. 
Conversion to incoming, $\radialFnSmat{-}$, or outgoing wave, 
$\radialFnSmat{+}$, boundary conditions appropriate for photoionization and 
recombination respectively is straightforward. 
\begin{eqnarray}
\radialFnSmat{\pm}=\sqrt{\frac{2}{\pi}}\radialFnKmat[\matrx{1} \mp i\matrx{K}]^{-1}.
\end{eqnarray}
\subsubsection{Expansion coefficients}

Eq. (\ref{eqn_expanded_scat}) allows us to write the radial wavefunction in
terms of the expansion coefficients and boundary amplitudes as follows,
\begin{equation}
\radialFnSmat{\pm}=\matrx{w}\AkCoeffMat{\pm},
\end{equation}
remembering that the R-matrix relates the radial function and its derivative 
we can write
\begin{eqnarray}
\matrx{w}\AkCoeffMat{\pm}&=&\matrx{R}\radialFnPSmat{\pm} \\
&=&\frac{1}{2} \matrx{w}  \left[ \matrx{E_k}-E \right]^{-1} \matrx{w}^T\radialFnPSmat{\pm} 
\end{eqnarray}
from which it is easy to see that
\begin{equation}
\matrx{A}^{\pm}=\frac{1}{2}\left[ \matrx{E_k}-E \right]^{-1}  \matrx{w}^T \matrx{F}'^{\pm}
\end{equation}
or alternatively
\begin{equation}
\matrx{A}^{\pm}= \frac{1}{2} \left[ \matrx{E_k}-E \right]^{-1}  \matrx{w}^T  \matrx{R}^{-1} \matrx{F}^{\pm}.
\end{equation}
To obtain the expansion coefficient all quantities need to be evaluated at a 
single radius.
However, following a standard scattering calculation we have the boundary 
amplitudes defined only at the R-matrix boundary, $a$, and the radial function 
defined only at the matching radius, $c$. 
R-matrix propagation is performed using the technique of 
Baluja~\textit{et al.} \cite{baluja82}
where a set of four matrices, $ \{ \matrx{\mathcal{R}}_{11},\matrx{\mathcal{R}}_{12},
\matrx{\mathcal{R}}_{21},\matrx{\mathcal{R}}_{22} \}$, are constructed which 
relate the R-matrix at $a$ to the R-matrix at $c$ as follows,
\begin{eqnarray}
\matrx{R}(c)&=&\matrx{\mathcal{R}}_{22}-\matrx{\mathcal{R}}_{21}[\matrx{\mathcal{R}}_{11}+\matrx{R}(a)]^{-1}\matrx{\mathcal{R}}_{12} \\
\matrx{R}(a)&=&\matrx{\mathcal{R}}_{12}[\matrx{\mathcal{R}}_{22}-\matrx{R}(c)]^{-1}\matrx{\mathcal{R}}_{21}-\matrx{\mathcal{R}}_{11}.
\end{eqnarray}
We also have the relation,
\begin{eqnarray}
 \matrx{F}(a)&=&\matrx{\mathcal{R}}_{12}\matrx{F}'(c)-\matrx{\mathcal{R}}_{11}\matrx{F}'(a)\\
 \matrx{F}(c)&=&\matrx{\mathcal{R}}_{22}\matrx{F}'(c)-\matrx{\mathcal{R}}_{21}\matrx{F}'(a),
\end{eqnarray}
which can be rearranged to give,
\begin{eqnarray}
\matrx{F}(a)&=&\matrx{\mathcal{R}}_{12}-\matrx{\mathcal{R}}_{11}\matrx{\mathcal{R}}_{21}^{-1}[\matrx{\mathcal{R}}_{22}-\matrx{R}(c)]\matrx{F}'(c)\\
\matrx{F}'(a)&=&\matrx{\mathcal{R}}_{21}^{-1}[\matrx{\mathcal{R}}_{22}\matrx{F}'(c)-\matrx{F}(c)].
\end{eqnarray}
This gives us the radial functions back propagated to the R-matrix boundary 
allowing construction of the wavefunction coefficients. 
We note that some care must be taken near channel thresholds where numerical 
instability can occur. Generally this can be avoided by increasing the 
forward propagation distance and, if necessary, increasing the numerical 
precision under which the linear algebra is performed. Expansion coefficients 
are calculated by a new outer region routine, COMPAK.
\subsection{Bound states and inner region dipoles}

Bound states required to describe the initial(/final) state for 
photoionization(/recombination) can be produced in several ways: 
they can be constructed using standard quantum chemistry techniques, with the 
proviso that the same set of orbitals must be used as was used for the 
target calculation;
they can be constructed from the inner region wavefunctions by considering all 
channels to be closed, using the outer region module BOUND \cite{sarpal91}; 
lastly, it can be a good approximation to take the lowest energy inner region 
wavefunction of the appropriate symmetry to represent the ground state of the 
neutral molecule \cite{tashiro10}.

Transition dipoles between inner region wavefunctions are calculated using 
a new, optimized and extended, version of DENPROP that can directly
use the close coupling basis of eq. (\ref{eqn_closecoup_inner}), 
CDENPROP (detailed elsewhere, \cite{harvey13}).

\section{CO$_2$: Models}
\subsection{Target}

The first step in a R-matrix calculation is construction of the target states, 
in the case of CO$_2$ photoionization, states of CO$_2^+$ at the equilibrium bond 
length of the neutral (for previous work on scattering from CO$_2^+$ using 
UKRmol see \cite{harvey07,harvey09,harvey11}).
A set of molecular orbitals were constructed from an initial Gaussian basis set 
(cc-pVTZ) \cite{dunning89} using a state averaged CASSCF \cite{knowles85,werner85}  
procedure with the quantum chemistry package Molpro \cite{werner12}. 
21 states of the ion and the ground state of the neutral were included in the 
averaging.
Choice of orbital set can have a strong influence on shape resonance features; 
in the spirit of the frozen core Hartree-Fock approximation, we chose the 
state averaging to be predominantly weighted towards the neutral with a small
(10\%) component of ionic states to improve the description of the target.

We note here that UKRmol is restricted to the use of abelian point groups, 
therefore D$_{2h}$ is the highest symmetry that can be used for CO$_2$. 
Where target states are degenerate we include both states (corresponding to two 
different irreducible representations in D$_{2h}$), this also has consequences 
for the treatment of the continuum, as a basis of real (tesseral) spherical 
harmonics, the appropriate angular basis for D$_{2h}$ and its subgroups, is used 
instead of the usual complex spherical harmonics.
This leads to some necessary modifications of eq. 
(\ref{eqn_rmat_photodipole_angular_momentum}-\ref{eqn_partial_wave_dipole_lab}), 
which are detailed in the appendix.
We looked at three different target models of increasing complexity 
(see table \ref{TargetModels}). 
All models have 6 electrons frozen in the core orbitals 
$(1-2\sigma_g,1\sigma_u)$.
Model 1 represents the target states with a single configuration state function
(CSF) and thus does not include electronic correlation in the ion. 
Model 2 includes single and double excitations into the valence orbitals not 
included in model 1. 
Model 3 consists of the full valence complete active space minus the $4\sigma_u$ 
orbital.
\begin{table}
\caption{\label{TargetModels}Target models: The bracket exponents denote the 
number of electrons placed in the set of orbitals contained in the bracket. The 
configurations column lists the maximum number of configurations used to 
represent a target state in that model (there is a small symmetry dependence 
to the number of configurations). }
\begin{indented}
\item[]\begin{tabular}{@{}lll}
\br
Model& Active space & CSFs \\
\mr
Model 1& $(1\mhyphen2\sigma_g,1\sigma_u)^6 (3\mhyphen4\sigma_g,2\mhyphen3\sigma_u,1\pi_u 1\pi_g)^{15}$                     & 1    \\
Model 2& $(1\mhyphen2\sigma_g,1\sigma_u)^6 (3\mhyphen4\sigma_g,2\mhyphen3\sigma_u,1\pi_u 1\pi_g)^{15}$                     & 440  \\
       & $(1\mhyphen2\sigma_g,1\sigma_u)^6 (3\mhyphen4\sigma_g,2\mhyphen3\sigma_u,1\pi_u 1\pi_g)^{14}(5\sigma_g,2\pi_u,4\sigma_u)^1$ &      \\
       & $(1\mhyphen2\sigma_g,1\sigma_u)^6 (3\mhyphen4\sigma_g,2\mhyphen3\sigma_u,1\pi_u 1\pi_g)^{13}(5\sigma_g,2\pi_u,4\sigma_u)^2$ &      \\
Model 3& $(1\mhyphen2\sigma_g,1\sigma_u)^6 (3\mhyphen5\sigma_g,2\mhyphen3\sigma_u,1\mhyphen2\pi_u 1\pi_g)^{15}$            & 3692 \\
\br
\end{tabular}
\end{indented}
\end{table}
\begin{table}
\caption{\label{TargetEnergies}Target energies (eV) relative to the 
groundstate for the various models, in comparison to experiment 
\cite{kelkensberg11}.
Bracketed figures: difference with experiment.}
\begin{indented}
\item[]\begin{tabular}{@{}lllll}
\br
State & Model 1 & Model 2 & Model 3 & Experiment \\
\mr
X$^2\Pi_g$       & 0.00        & 0.00        & 0.00        & 0.0\\
A$^2\Pi_u$       & 4.70 (0.90) & 4.12 (0.32) & 3.97 (0.17) & 3.8\\
B$^2\Sigma^{+}_u$& 5.33 (1.03) & 4.77 (0.47) & 4.45 (0.15) & 4.3\\
C$^2\Sigma^{+}_g$& 7.14 (1.54) & 5.97 (0.37) & 5.77 (0.17) & 5.6\\
\br
\end{tabular}
\end{indented}
\end{table}
Table \ref{TargetEnergies} shows the energies of the lowest 4 ionic states. 
Model 1 gives generally poor agreement with experimental energies, model 2, 
is significantly better, and model 3 agrees to within $~$0.2~eV.

The lowest inner region wavefunction of the appropriate symmetry was 
used as the ground state of the neutral.

\subsection{Inner region}
Continuum orbitals up to $l=5$ and spanning an energy range from 0 to 
3.5~Hartree were generated by optimizing a set of GTO to represent Coulomb 
functions. 3 virtual orbitals were included in each symmetry 
to improve the description of inner region polarization. Single channel 
calculations were performed using the model 1 description of the ion and 
multichannel calculations with models 2 and 3. With model 2 and 3 we included 96 
states in the close coupling expansion. 
  
\subsection{Outer region}

We matched to Coulomb functions at the R-matrix boundary: for molecules, such as 
CO$_2$, with no permanent dipole, this approximation works well, at least at the
level of the background cross section and shape resonances, and offers 
significant computational saving when there are many channels. 
Narrow resonance features however, are sensitive to channel coupling in the 
outer region and we would not expect to get good positions and widths without 
outer region propagation. 

\section{CO$_2$: Results}

\subsection{Partial cross sections}

\begin{figure}
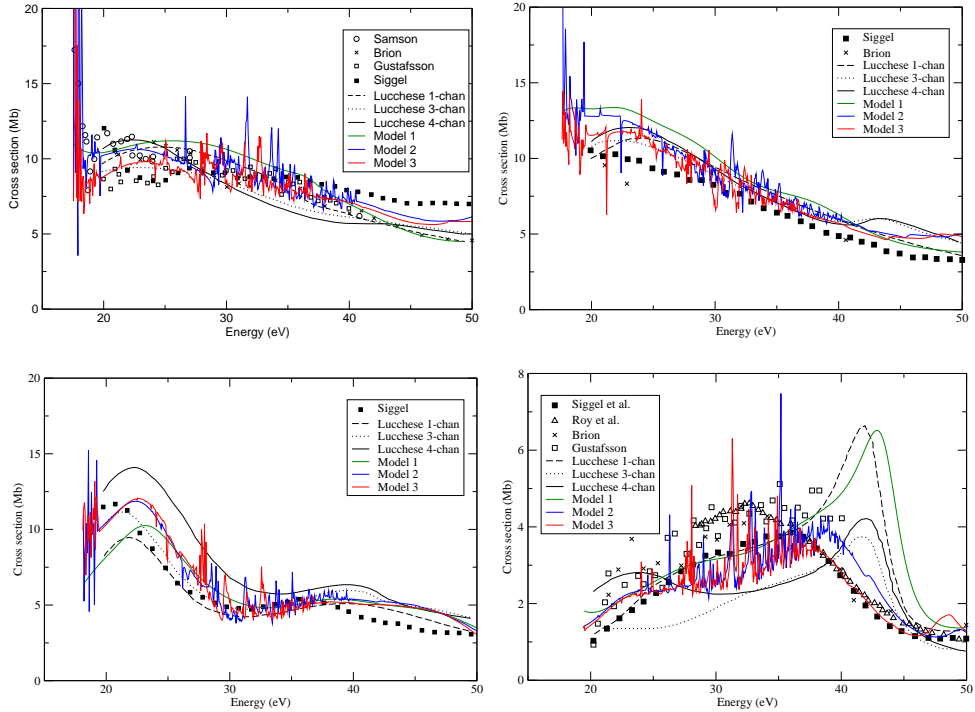

\centering
\subfloat{\includegraphics[width=0.48\textwidth]{./figs/pxsec.X.eps}  }%
\subfloat{ \includegraphics[width=0.48\textwidth]{./figs/pxsec.A.eps}  }\\
\subfloat{ \includegraphics[width=0.48\textwidth]{./figs/pxsec.B.eps}  }%
\subfloat{ \includegraphics[width=0.48\textwidth]{./figs/pxsec.C.eps}  }
 \caption{ Partial cross sections: Top left, final ion state X$^2\Pi_g$. Top right, 
 A$^2\Pi_u$. Bottom left, B$^2\Sigma^{+}_u$. Bottom right, C$^2\Sigma^{+}_g$. 
 Experimental results: Samson et al. \cite{samson73},  Gustaffson et al. \cite{gustafsson78}, 
 Siggel et al. \cite{siggel93}, Brion and Tan \cite{brion78} and Roy et al. \cite{roy84}.
 Previous thoeretical results, Lucchese et al. \cite{lucchese90}} 
\label{figPartialXsecs}
\end{figure}

Fig. (\ref{figPartialXsecs}) shows orientationally averaged partial 
cross sections leaving the ion in the ground and first three excited 
states. 
The first feature to note is the presence of a high number of narrow 
resonance features in the R-matrix cross sections, these are autoionizing 
resonances associated with the various excitation thresholds included in the 
models. 
The single channel cross sections, with no excitation, are smooth. 
The experimental results we compare to here are of insufficient resolution to 
resolve the resonances, and the accurate characterization of such resonances, 
traditionally a strength of R-matrix approaches, has been left for future work. 
In this work we concern ourselves with the background cross sections and broader 
shape resonance features.

The agreement with experiment and previous theory is excellent,
with the CI models (2 and 3) generally in better agreement, up to $45$~eV, than model 1.
Above $45$~eV the CI models display some unphysical pseudoresonances related to
the omission of highly excited states in the close coupling expansion that are 
implicitly included in the second summation of eq. (\ref{eqn_closecoup_inner}).  
Model 1, the single channel, static exchange model, gives slightly worse 
resonance positions compared to the previous work \cite{lucchese90} (labelled Lucchese: 1 chan 
in the figures) at a similar level of approximation performed using the 
Schwinger variational approach. 
This difference is primarily due to the choice of orbitals (neutral HF vs state 
averaged CASSCF over both neutral and ionic states); our orbital choice 
includes some degree of orbital relaxation which tends to shift resonance 
positions to higher energy. 
Another difference is in the choice of gauge, our work uses the length gauge, 
as opposed to a mixed gauge approach in the previous theoretical work.

\begin{figure}
\centering

\subfloat{\includegraphics[width=0.96\textwidth]{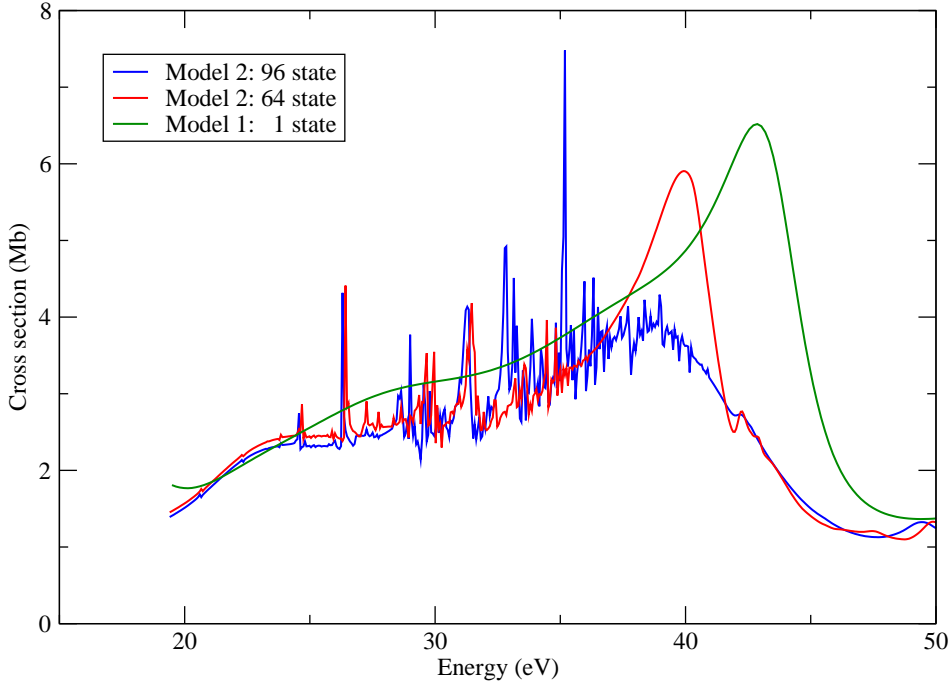}  }%

\caption{ Partial cross sections: Final ion state  C$^2\Sigma^{+}_g$. 
 Comparison of models with different number of ionic states included.} 
\label{figC-PartialXsecCompare}
\end{figure}

Previous theoretical work found a high, narrow shape resonance in the C channel at 
around $42$~eV, approximately $5$~eV above the IP of the ion, that was not evident in 
experimental cross sections. 
Various attempts to reconcile theory and experiment were made at the time, 
including vibrational averaging \cite{lucches82a}, and the inclusion of channel 
coupling and initial state correlation \cite{lucchese82,lucchese90}, with 
partial success.
It was speculated that the discrepancy was due to the need for correlation in the 
ion and many more excited states of the ion to be included in the 
channel coupling.
Model 2 and 3 include both these and give very good agreement to experiment.
In Fig  (\ref{figC-PartialXsecCompare}) we see that the number of ionic 
channels plays a strong role in suppression of the resonance. 
With 64 states included the resonance is somewhat lowered in amplitude and 
significantly shifted in position, going to 96 states only has a small 
effect on position, but a dramatic effect on resonance height, which we 
attribute to loss of flux into highly excited ionic states.
    
%

Finally we note that cross sections near to and beyond the IP of the ion must be 
treated with some caution; for reliable treatment of the intermediate energy range 
(from close to the ionization threshold of the target up to several times this threshold), 
it has been found to be important in the accurate calculation of scattering 
and photoionization observables to account for highly excited electronic 
states of the target and the target continuum that are not included in the 
standard close-coupling approach. 
An approach that has found a great deal of success in atomic photoionization and 
scattering calculations and is beginning to be applied  to the molecular case 
is the R-matrix with pseudostates method (RMPS) \cite{gorfinkiel05}. 
As well as a rigorous treatment of the intermediate energy regime, RMPS has 
the benefit of converging the polarizability of the ion, which can have a 
strong effect on resonance position. 

%

\subsection{Photoelectron angular distributions}

Two common experimental alignment distributions are: aligned with with the photon 
polarization and planar delocalized perpendicular (anti-aligned) to the 
photon polarization. These arise in the impulsive laser alignment of molecules, 
where a rotational wavepacket is produced that cycles between alignment, 
anti-alignment and random alignment \cite{friedrich95,kumar96,seidman99,stapelfeldt03,zeng09}. 
In this section we present photo-electron angular distributions (PAD) for these 
two scenarios. 

\begin{figure}
\centering
\subfloat{\includegraphics[width=62mm]{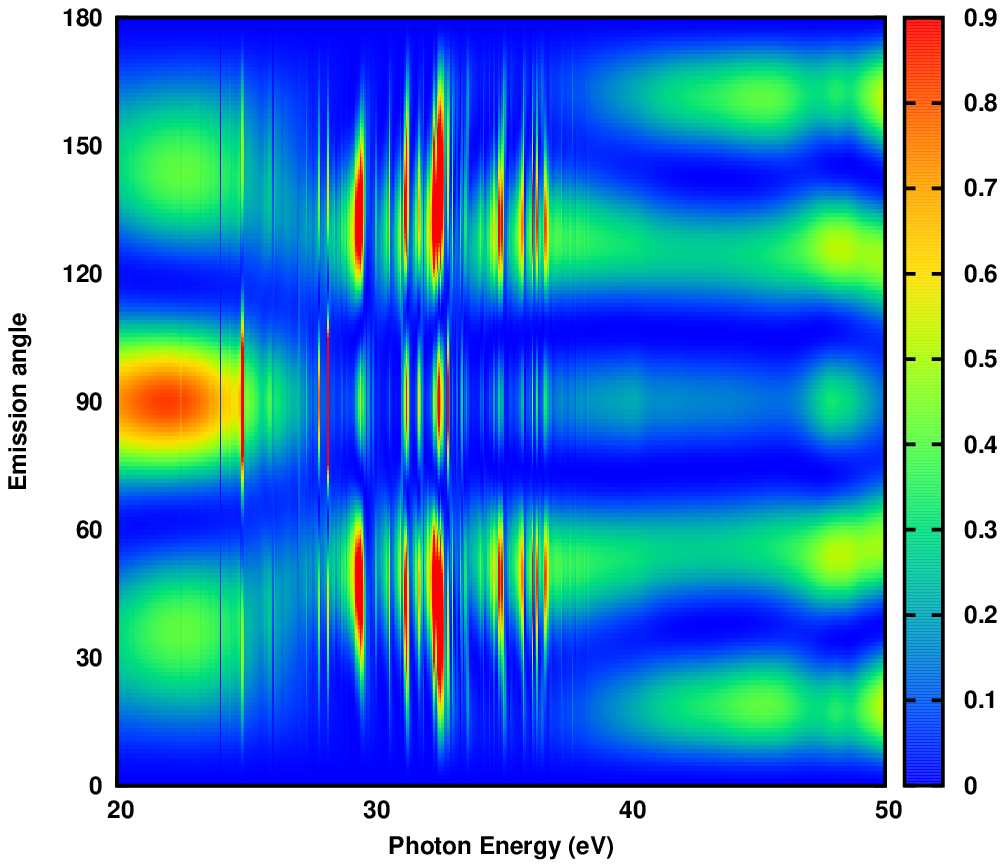}}%
\quad
\subfloat{\includegraphics[width=62mm]{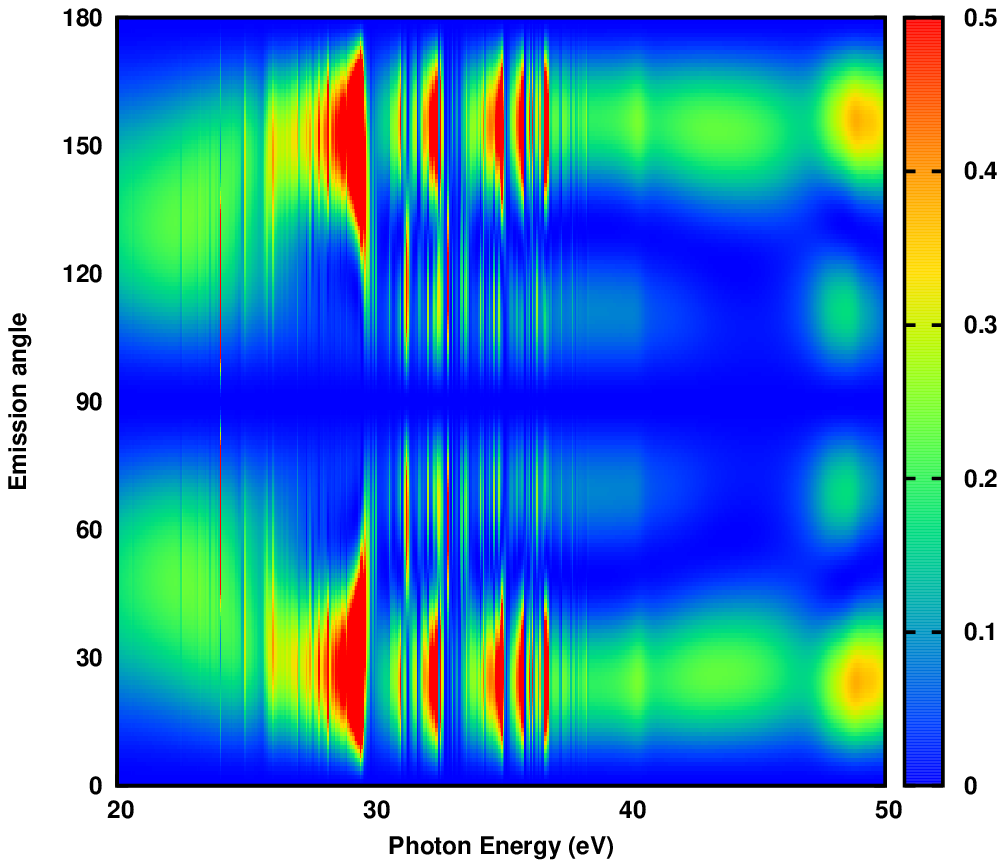}}\\
\subfloat{\includegraphics[width=62mm]{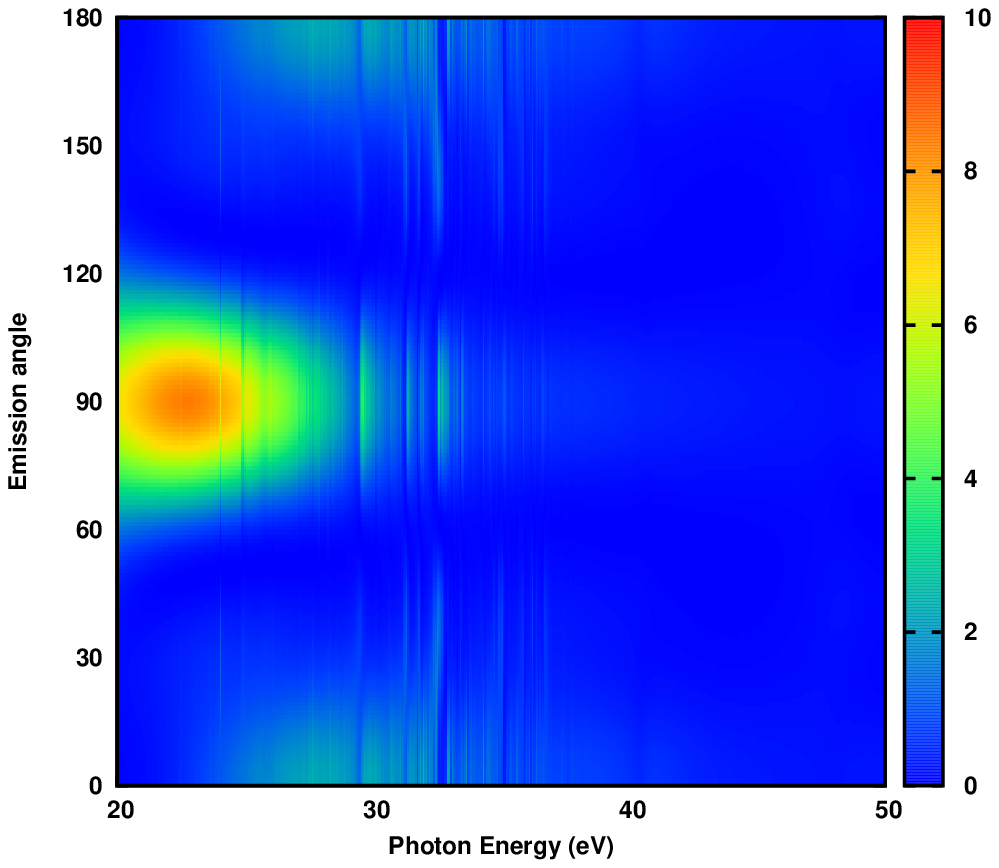}}%
\quad
\subfloat{\includegraphics[width=62mm]{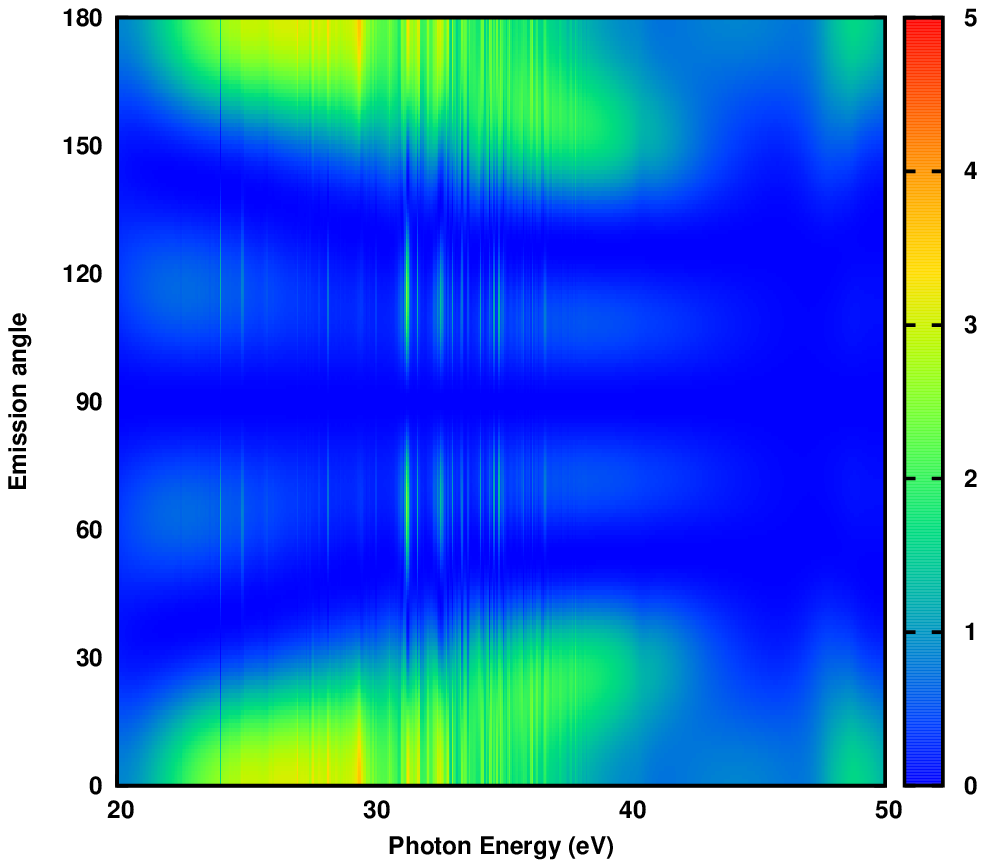}}
 \caption{Aligned PAD, the emission angle is defined relative to the lab 
 frame photon polarization (which defines the lab z-axis), the magnitude has 
 units Mb: Top left, X$^2\Pi_g$. Top right, A$^2\Pi_u$. Bottom left, 
 B$^2\Sigma^{+}_u$. Bottom right, C$^2\Sigma^{+}_g$} 
\label{figApad}
\end{figure}

\begin{figure}
\centering
\subfloat{\includegraphics[width=62mm]{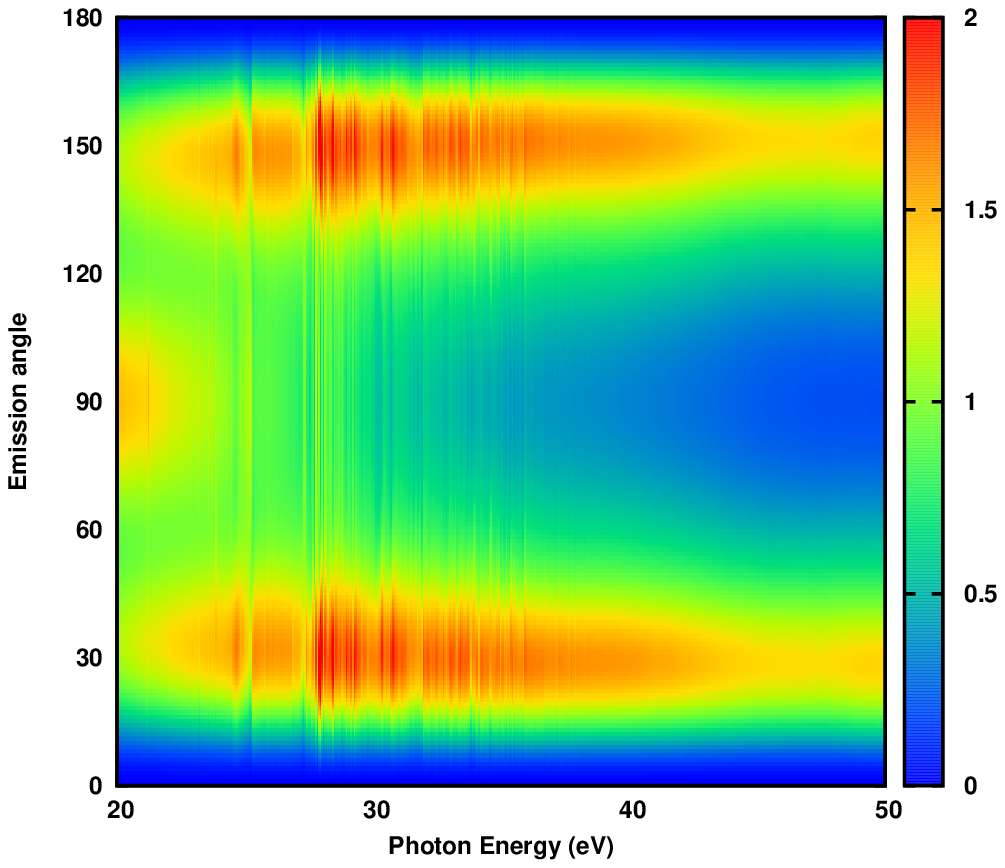}}%
\quad
\subfloat{\includegraphics[width=62mm]{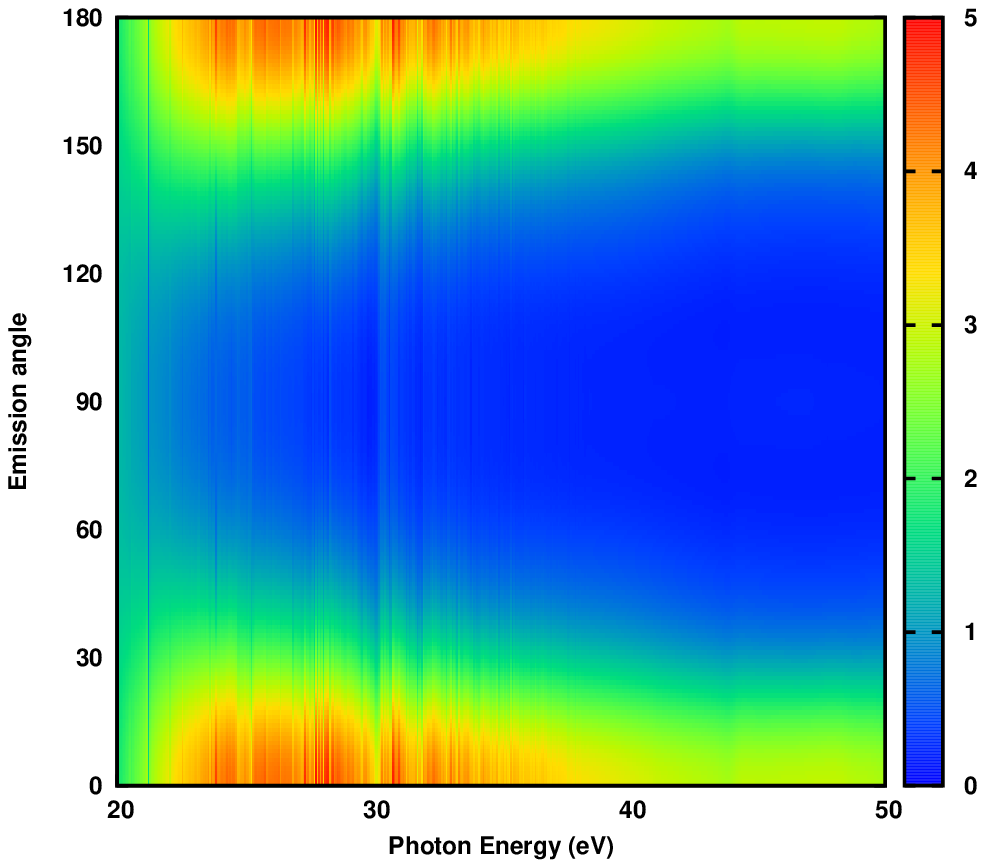}}\\
\subfloat{\includegraphics[width=62mm]{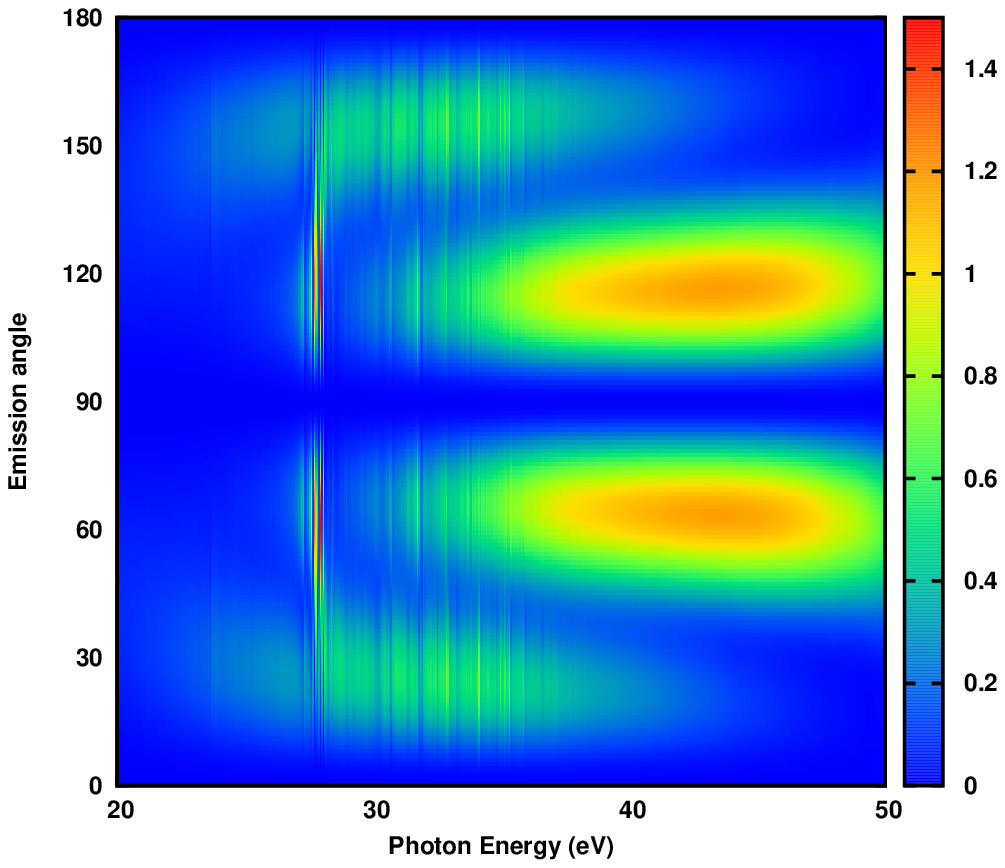}}%
\quad
\subfloat{\includegraphics[width=62mm]{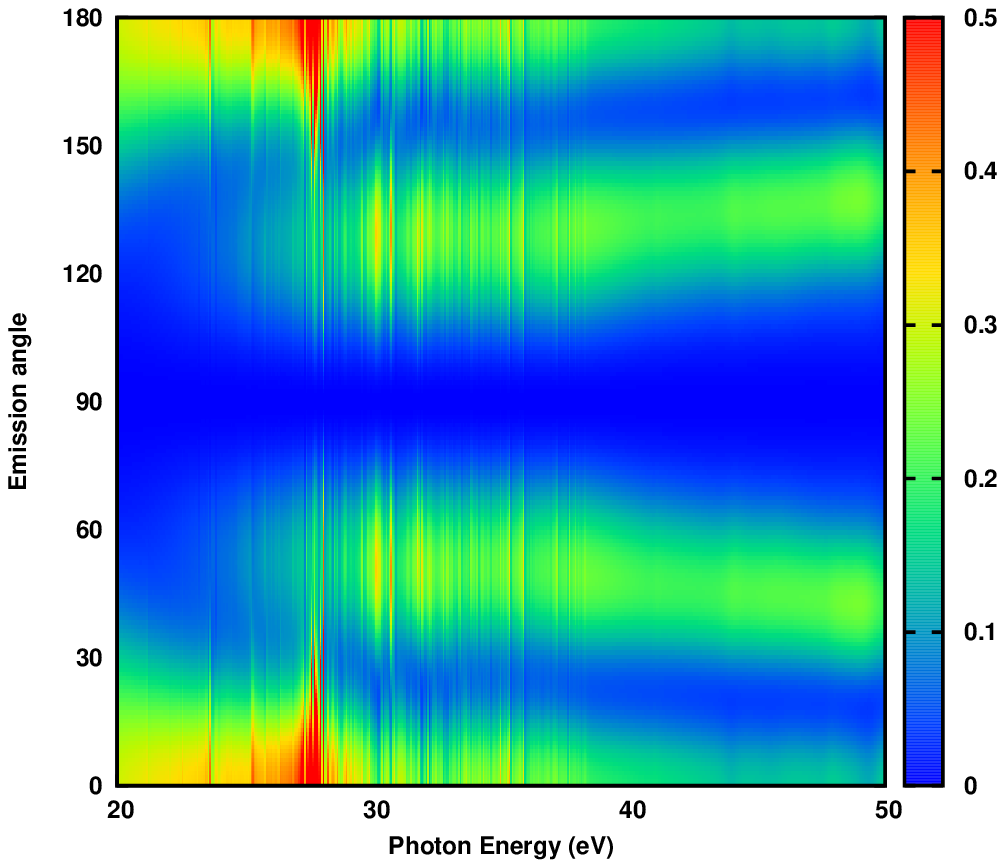}}
 \caption{ Anti-aligned PAD, the emission angle is defined relative to the lab 
 frame photon polarization (which defines the lab z-axis), the magnitude has 
 units Mb: Top left, X$^2\Pi_g$. Top right, A$^2\Pi_u$. Bottom left, 
 B$^2\Sigma^{+}_u$. Bottom right, C$^2\Sigma^{+}_g$} 
\label{figAApad}
\end{figure}

Figures \ref{figApad} and \ref{figAApad} show the $p_z-p_x$ emission plane for 
aligned and anti-aligned molecules respectively. 
The lab $z-$axis is defined by the (linearly polarized) photon polarization. 
In both these cases the cylindrical symmetry of the system is preserved 
leading to a cylindrically symmetric (around the $z-$axis) PAD. 
We therefore lose no information in looking at a 2D momentum cut.

We see that both aligned and anti-aligned results have rich angular structure, 
one useful method for the interpretation of angular distributions is to 
consider the angular momentum of the ejected photo-electron, the angular pattern 
in a particular energy region can be dominated by a particular partial wave, 
such as when there is a resonance in that partial wave, and interference 
between partial waves is a sensitive probe of the photoelectron-ion potential
\cite{reid03}.

The ground state of the neutral has  $^1\Sigma_g$ symmetry, for the aligned case 
the dipole operator has $\sigma_u$ symmetry leading to final state symmetry 
of $^1\Sigma_u$. This leads to a $\pi_u$, $\pi_g$, $\sigma_g$ and $\sigma_u$ 
symmetry of the continuum in the X, A, B and C channels respectively.  
For the anti-aligned case the dipole operator has $\pi_u$ symmetry leading to 
final state symmetry of $^1\Pi_u$. This leads to a $\sigma_u$, $\sigma_g$, 
$\pi_g$ and $\pi_u$, symmetry of the continuum in the X, A, B and C channels 
respectively.  $\sigma$ and $\pi$ continua correspond to partial waves with 
$|m|=0$ and $|m|=1$ respectively. l is even or odd (gerade or ungerade) and a 
particular partial wave has $|m|$ longitudinal nodes $l-|m|$ latitudinal nodes.   
Taking the X channel in the aligned case as an example we can see that it is 
dominated by the $l=3$ partial wave, with a smaller contribution from $l=1$, between $20$ and $30$~eV. 
At higher energies, from $40$~eV the $l=5$ partial wave becomes dominant.
 
\section{Conclusions}

We have presented extensions to the polyatomic UKRmol electron-molecule 
scattering codes that allow calculation of photoionization and recombination, 
from both oriented and orientationally averaged molecules. 
Our new codes have been applied to CO$_2$, with good agreement to experiment and 
alternative theoretical techniques for orientationally averaged quantities.
The inclusion of both correlation, and channel coupling between 
many highly excited states of the ion was found to be important for the accurate 
description of the shape resonance in the C$^2\Sigma^{+}_g$ partial photoionization 
cross section. 
We note that our angularly resolved calculations have been used in two joint 
theoretical-experimental studies of aligned CO$_2$, the first measures 
photoelectron angular distributions from aligned CO$_2$ using a HHG photon 
source \cite{rouzee13}, the second measured HHG from aligned CO$_2$ (Harvey et 
al. in prep.).
In both cases good agreement between theory and experiment was achieved. 
We anticipate that, as has been the case for atoms, the R-matrix approach 
to photoionization will be a fruitful method for the accurate study of 
photoionization and recombination processes in polyatomic molecules.  
 
\section*{Appendix}
\subsection*{The photoionization dipole in D$_{2h}$}

UKRmol uses real spherical harmonics in the partial wave expansion of the 
continuum, eq.~(\ref{eqn_rmat_photodipole_angular_momentum}), and to represent 
the dipole operator when calculating dipoles between molecular orbitals, 
eq.~(\ref{eqn_dipole_operator}). The lab frame dipole in this basis becomes 
(dropping the spin related Clebsch-Gordan coefficient for sake of 
clarity),
\begin{eqnarray}\label{eqn_partial_wave_dipole_lab_d2h}
\boldsymbol{d}^{(\mathrm{Re})'}_{fi}(\kfinalLAB)=\sum_{l_f m_f' m_f}i^{-l_f}{e^{i\coulombphase{l_f}}}\teharm{l_f}{m_f'}(\kdirLAB)\rerotmat{l}{m_f'}{m_f} \boldsymbol{d}^{(\mathrm{Re})}_{fl_fm_f,i}(E)\MTXrerotmat{1T},
\end{eqnarray} 
where $\teharm{l_f}{m_f'}(\kdirLAB)$ are the real spherical harmonics and 
$\rerotmat{l}{m_f'}{m_f}(\euler)$ are the rotation matrices in the basis of real 
spherical harmonics \cite{blanco97}. For the case of linear molecules, starting in the 
molecular frame,
\begin{eqnarray}\label{eqn_partial_wave_dipole_mol_d2h}
\boldsymbol{d}^{(\mathrm{Re})}_{fi}(\kfinalMOL)=\sum_{l_f m_f}i^{-l_f}{e^{i\coulombphase{l_f}}}\teharm{l_f}{m_f}(\kdirMOL) \boldsymbol{d}^{(\mathrm{Re})}_{fl_fm_f,i}(E),
\end{eqnarray} 
and using the unitary transformation between the real and complex spherical 
harmonics \cite{blanco97},
\begin{eqnarray}
\boldsymbol{S}_{l_f}=\matrx{C}^{l_f}\boldsymbol{Y}_{l_f}
\end{eqnarray} 
where the spherical harmonics have been written in vector form, we get,
\begin{eqnarray}\label{eqn_partial_wave_dipole_mol_d2h_to_dinfh}
\boldsymbol{d}_{fi}(\kfinalMOL)=\sum_{l_f m_f}i^{-l_f}{e^{i\coulombphase{l_f}}}\spharm{l_f}{m_f}(\kdirMOL) \sum_{m_f'}\matrx{C}^{1\dagger} \boldsymbol{d}^{(\mathrm{Re})}_{fl_fm_f',i}(E){C}^{l_f}_{m_f'm_f},
\end{eqnarray} 
and identifying,
\begin{eqnarray}
\boldsymbol{d}_{fl_fm_f,i}(E)=\sum_{m_f'}\matrx{C}^{1\dagger} \boldsymbol{d}^{(\mathrm{Re})}_{fl_fm_f',i}(E){C}^{l_f}_{m_f'm_f},
\end{eqnarray} 
we recover the molecular frame dipole in the partial wave basis and can 
transform to the lab frame as before.

\section*{Acknowledgements}
The authors would like to acknowledge useful discussions with 
Jonathan Tennyson. 
We acknowledge the support of Einstein foundation project A-211-55 Attosecond 
Electron Dynamics. 

\section*{References}

\bibliographystyle{unsrt}
\bibliography{photo-rmat-paper2,sub-prep-arxiv}

\end{document}